\documentclass{article}

\PassOptionsToPackage{numbers, compress}{natbib}

\usepackage[preprint]{neurips_2024}

\usepackage[utf8]{inputenc} % allow utf-8 input
\usepackage[T1]{fontenc}    % use 8-bit T1 fonts
\usepackage{hyperref}       % hyperlinks
\usepackage{url}            % simple URL typesetting
\usepackage{booktabs}       % professional-quality tables
\usepackage{amsfonts}       % blackboard math symbols
\usepackage{nicefrac}       % compact symbols for 1/2, etc.
\usepackage{microtype}      % microtypography
\usepackage{xcolor}         % colors
\usepackage{authblk}
\usepackage{subcaption}
\usepackage{graphicx}
\usepackage{xspace}
\usepackage{amsmath}
\usepackage{ulem}
\usepackage{algorithm}
\usepackage{algorithmic}
\usepackage{multirow}
\usepackage{multicol}
\usepackage{adjustbox}
\usepackage{wrapfig}
\usepackage{array}
\usepackage{booktabs}
\usepackage{listings}

\definecolor{codegreen}{rgb}{0,0.6,0}
\definecolor{codegray}{rgb}{0.5,0.5,0.5}
\definecolor{codepurple}{rgb}{0.58,0,0.82}
\definecolor{backcolour}{rgb}{0.95,0.95,0.92}

\lstdefinelanguage{markdown}{
    morekeywords={\#, \#\#, \#\#\#, \#\#\#\#, -, *, >, `, ```}, 
    sensitive=true, 
    morecomment=[l]{//}, 
    morecomment=[s]{/*}{*/}, 
    morestring=[b]", 
    morestring=[b]', 
}

\lstdefinestyle{markdownstyle}{
    language=markdown,
    basicstyle=\ttfamily\footnotesize,
    keywordstyle=\color{blue}\bfseries,
    commentstyle=\color{gray}\ttfamily,
    stringstyle=\color{red}\ttfamily,
    showstringspaces=false,
    breaklines=true,
    frame=single,
    backgroundcolor=\color{gray!10}
}
\lstdefinestyle{mystyle}{
    backgroundcolor=\color{backcolour},   
    commentstyle=\color{codegreen},
    keywordstyle=\color{black},
    numberstyle=\tiny\color{black},
    stringstyle=\color{codegreen},
    basicstyle=\ttfamily\footnotesize,
    breakatwhitespace=false,         
    breaklines=true,                 
    captionpos=b,                    
    keepspaces=true,                                   
    numbersep=5pt,                  
    showspaces=false,                
    showstringspaces=false,
    showtabs=false,                  
    tabsize=2,
    language=Python
}

\newcommand{\ourmethod}{AutoSurvey\xspace}

\title{AutoSurvey: Large Language Models Can Automatically Write Surveys}

\author{%
  \textbf{Yidong Wang}$^{1,2}$\thanks{Equal contribution. yidongwang37@gmail.com, qguo@smail.nju.edu.cn; Yidong Wang did this work during his internship at Squirrel AI.} , \textbf{Qi Guo}$^{2,3*}$, \\ \textbf{Wenjin Yao}$^{2}$, \textbf{Hongbo Zhang}$^{1}$, \textbf{Xin Zhang}$^{4}$, \textbf{Zhen Wu}$^{3}$,\textbf{Meishan Zhang}$^{4}$, \\ \textbf{Xinyu Dai}$^{3}$, \textbf{Min Zhang}$^{4}$, \textbf{Qingsong Wen}$^{5}$, \textbf{Wei Ye}$^{2}$\thanks{Correspondence to: wye@pku.edu.cn, zhangsk@pku.edu.cn, zhangyue@westlake.edu.cn.} , \textbf{Shikun Zhang}$^{2\dagger}$, \textbf{Yue Zhang}$^{1\dagger}$
}
\affil{\small{$^{1}$Westlake University, $^{2}$Peking University,\\
$^{3}$Nanjing University, $^{4}$Harbin Institute of Technology,
Shenzhen, $^{5}$Squirrel AI}
}

\begin{document}

\maketitle

\begin{abstract}
  This paper introduces \ourmethod, a speedy and well-organized methodology for automating the creation of comprehensive literature surveys in rapidly evolving fields like artificial intelligence. Traditional survey paper creation faces challenges due to the vast volume and complexity of information, prompting the need for efficient survey methods. While large language models (LLMs) offer promise in automating this process, challenges such as context window limitations, parametric knowledge constraints, and the lack of evaluation benchmarks remain. \ourmethod addresses these challenges through a systematic approach that involves initial retrieval and outline generation, subsection drafting by specialized LLMs, integration and refinement, and rigorous evaluation and iteration. Our contributions include a comprehensive solution to the survey problem, a reliable evaluation method, and experimental validation demonstrating \ourmethod's effectiveness. We open our resources at \url{https://github.com/AutoSurveys/AutoSurvey}.
\end{abstract}

\section{Introduction}

\begin{figure}[h!]
\centering
\begin{tabular}{cc}
\begin{minipage}{0.25\textwidth}
    \centering
    \begin{subfigure}{\textwidth}
        \centering
        \includegraphics[width=0.98\textwidth]{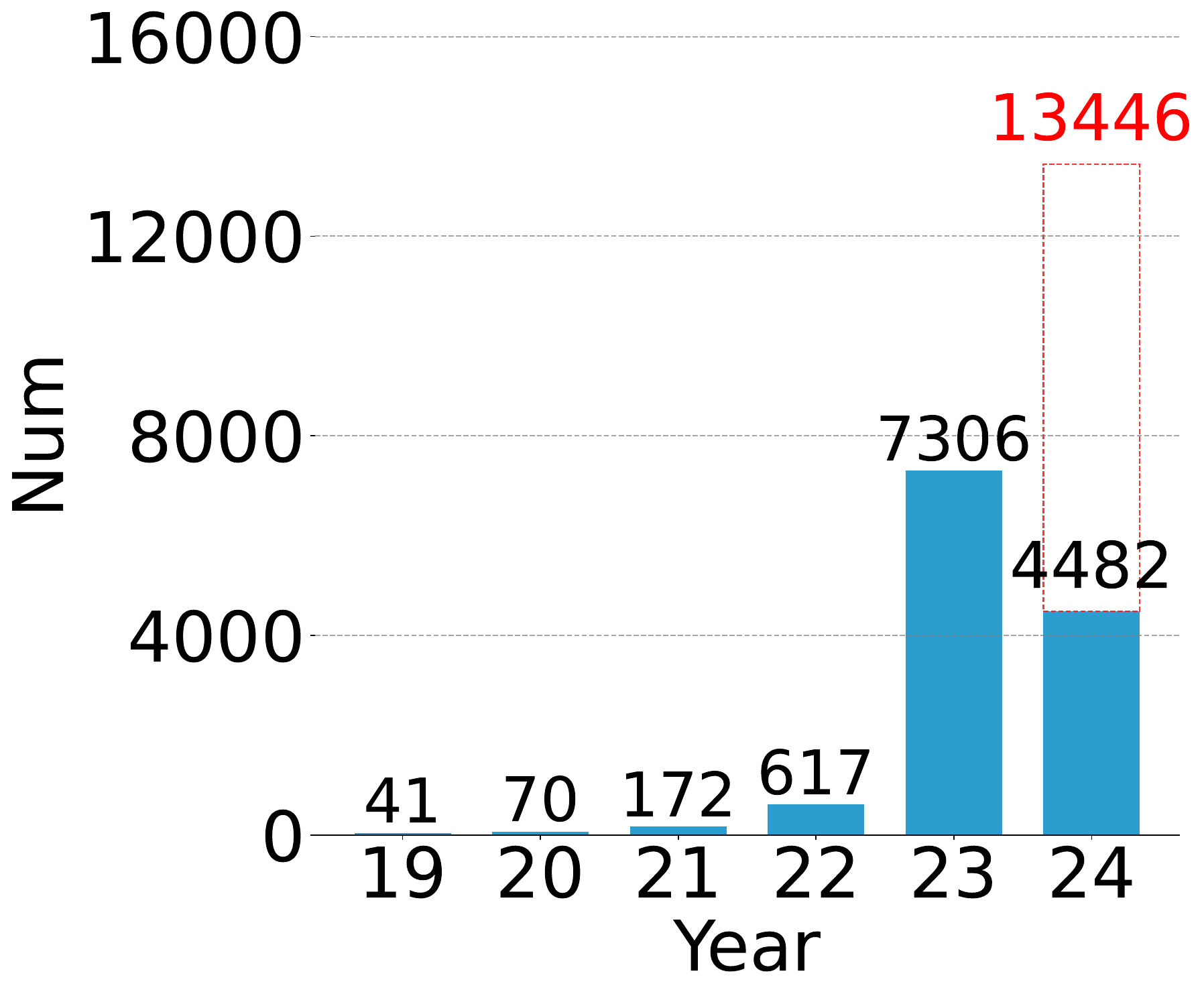}
        \caption{\#Paper about LLM.}
        \label{fig-paper-per-year}
    \end{subfigure} \\
    \begin{subfigure}{\textwidth}
        \centering
        \includegraphics[width=0.98\textwidth]{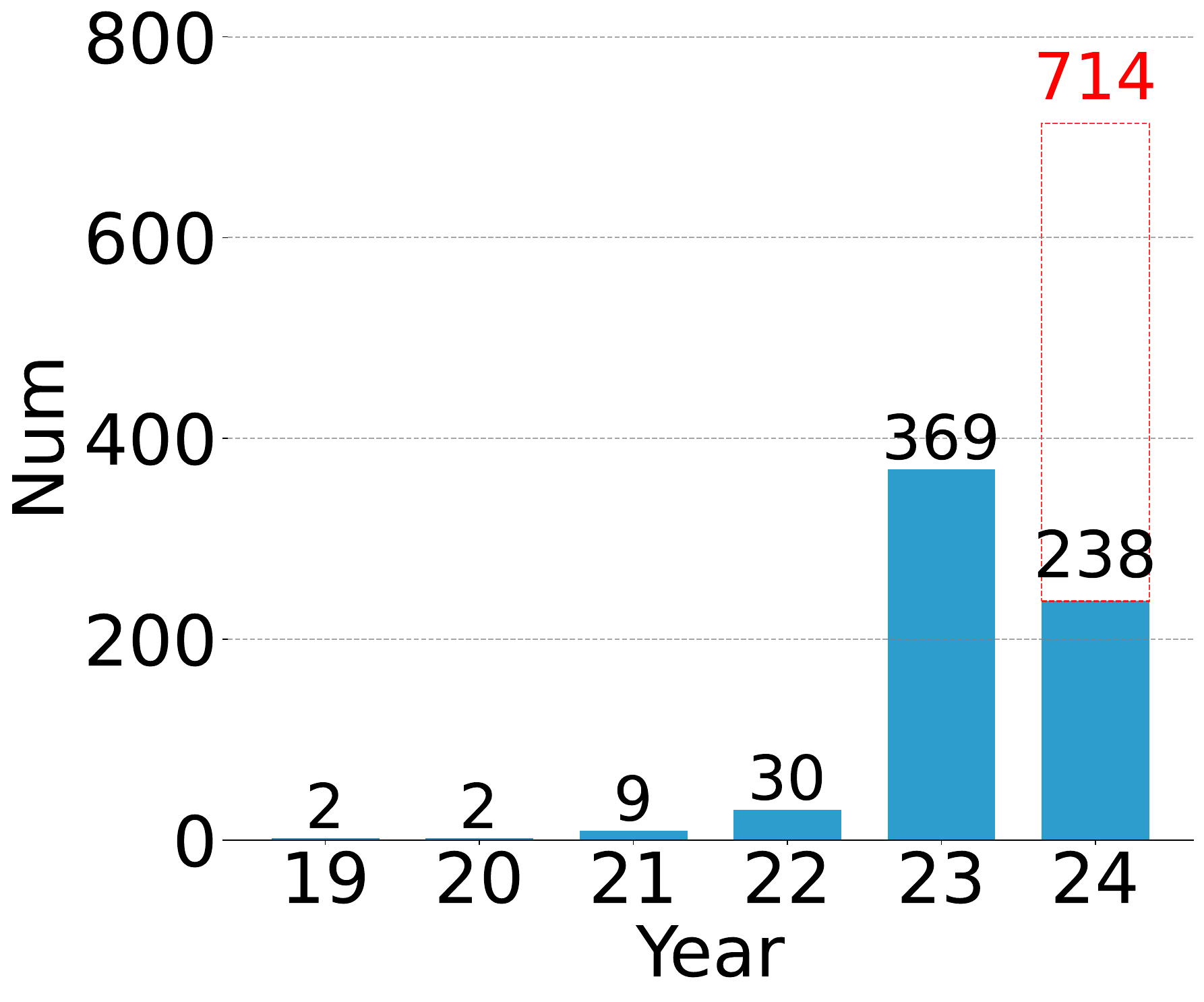}
        \caption{\#Survey about LLM.}
        \label{fig-survey-per-year}
    \end{subfigure}
\end{minipage} &
\begin{minipage}{0.55\textwidth}
    \centering
    \begin{subfigure}{\textwidth}
        \centering
        \includegraphics[width=0.98\textwidth]{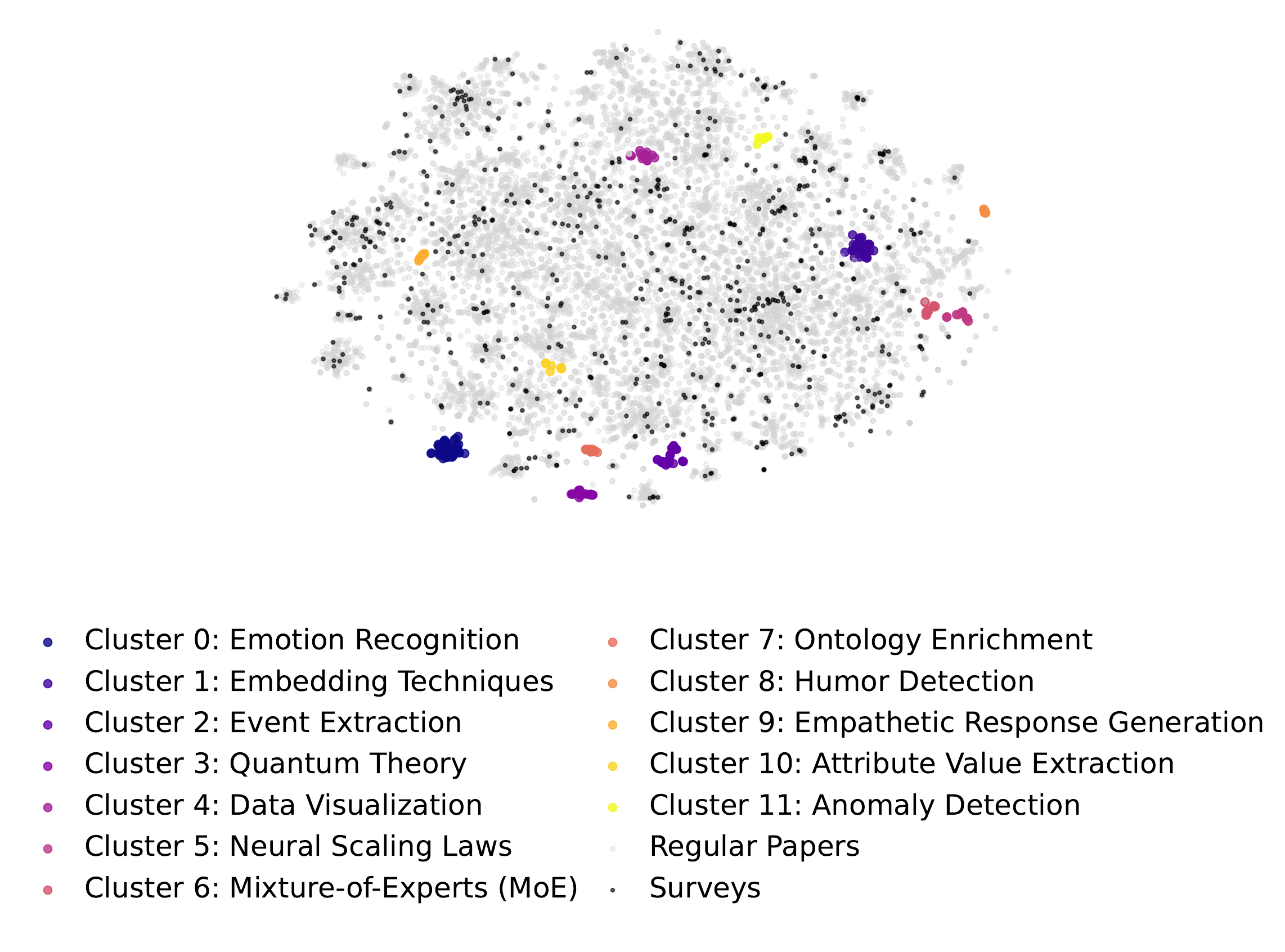}
        \caption{T-SNE visualization of surveys and papers about LLMs. Clusters represent groups of papers identified through clustering, which currently lack comprehensive survey coverage.}
        \label{fig-survey-tsne}
    \end{subfigure}
\end{minipage}
\end{tabular}
\caption{Depicting growth trends from 2019 to 2024 in the number of LLMs-related papers (a) and surveys (b) on arXiv, accompanied by a T-SNE visualization. The data for 2024 is up to April, with a red bar representing the forecasted numbers for the entire year. While the number of surveys is increasing rapidly, the visualization reveals areas where comprehensive surveys are still lacking, despite the overall growth in survey numbers. The research topics of the clusters in the T-SNE plot are generated using GPT-4 to describe their primary focus areas. 
\textbf{These clusters of research voids can be addressed using \ourmethod at a cost of \$1.2 (cost analysis in Appendix \ref{appendix:computational analysis}) and 3 minutes per survey. An example survey focused on Emotion Recognition using LLMs is in Appendix \ref{appendix:survey_example}.}
}
\label{fig-survey-paper-demo}
\end{figure}

Survey papers provide essential academic resources, offering comprehensive overviews of recent research developments, highlighting ongoing trends, and identifying future directions~\cite{pouyanfar2018survey,chang2023survey,zhao2023survey,khan2022transformers}. However, crafting these surveys is increasingly challenging, especially in the fast-paced domain of Artificial Intelligence including large language models(LLMs)~\cite{lecun2015deep,goodfellow2016deep,achiam2023gpt,kirillov2023segment}. Figure~\ref{fig-paper-per-year} illustrates a significant trend: in just the first four months of 2024 alone, over 4,000 papers containing the phrase "Large Language Model" in their titles or abstracts were submitted to arXiv. This surge highlights a critical academic issue: the rapid accumulation of new information often outpaces the capacity for comprehensive scholarly review and synthesis, emphasizing the growing need for more efficient methods to synthesize the expanding literature. Moreover, as depicted in Figure~\ref{fig-survey-per-year}, while the number of survey papers has rapidly increased, the growing difficulty of producing traditional human-authored survey papers—due to the sheer volume and complexity of data—remains a significant challenge. This challenge is evidenced by the lack of comprehensive surveys in many fields (Figure~\ref{fig-survey-tsne}), which hinders knowledge transfer and makes it difficult for new researchers to efficiently navigate the vast amount of available information.

The advent of LLMs~\cite{achiam2023gpt,touvron2023llama} presents a promising avenue for addressing these challenges. These models, trained on extensive text corpora, demonstrate remarkable capabilities in understanding and generating human-like text, even in long-context scenarios~\cite{chen2023extending,chen2023longlora,wang2024augmenting}. Despite these advancements, the practical application of LLMs to survey generation is fraught with challenges. Firstly, \textbf{context window limitations}: LLMs encounter inherent restrictions in output length due to limited processing windows~\cite{liu2024lost,kaddour2023challenges,shi2023large,li2023long,li2024long}. While several advanced large models, including GPT-4 and Claude 3, support inputs exceeding 100k tokens, their output is still limited to fewer than 8k tokens (the output length of GPT-4 is 8k, and the output length of Claude 3 is 4k). Writing a comprehensive survey typically requires reading hundreds of papers, resulting in input sizes far beyond the capacity of even the most advanced models. Moreover, a well-written survey itself spans tens of thousands of tokens, making it highly challenging to generate such extensive content directly with large models.
 Secondly, \textbf{parametric knowledge constraints}: Sole reliance on an LLM's internal knowledge is insufficient for producing surveys that require comprehensive and accurate references~\cite{wang2023surveyfact,ji2023survey,shao2024assisting}. LLMs may generate content based on inaccuracies or even non-existent ``hallucinated'' references. Moreover, these models cannot incorporate the latest studies not included in their training data, which limits the breadth and depth of the surveys they generate. Thirdly, \textbf{the lack of evaluation benchmark:} after production, reliable metrics to evaluate the quality of outputs from LLMs are lacking. Relying on human review for quality assessment is not only resource-intensive but also lacks scalability~\cite{pandalm2024,zheng2024judging,yu2024kieval}. This presents a significant obstacle to the widespread adoption of LLMs for academic synthesis, where rigorous standards of accuracy and reliability are paramount.

In response to these challenges, we introduce \ourmethod: a speedy and well-organized methodology for conducting comprehensive literature surveys. Specifically, \ourmethod's primary innovations include: \textbf{logical parallel generation}: \ourmethod employs a two-stage generation approach to parallelly generate survey content efficiently. Initially, multiple LLMs work concurrently to create detailed outlines. A final, comprehensive outline is then synthesized from these individual outlines, setting a clear framework for content development. Subsequently, each subsection of the survey is generated in parallel and guided by the outline, which significantly accelerates the process. To overcome potential transition and consistency issues due to segmented generation phases, \ourmethod integrates a systematic revision phase. After the initial parallel generation, each section undergoes thorough revision and polishing, ensuring smooth transitions and enhanced overall document consistency. The sections are then seamlessly merged to produce a cohesive and well-organized final document. \textbf{Real-time knowledge update}: \ourmethod incorporates a Real-time Knowledge Update mechanism using a Retrieval-Augmented Generation (RAG) approach~\cite{gao2023retrieval,lewis2020retrieval,jiang2023active}. This feature ensures that every aspect of the survey reflects the most current studies. When a survey topic is input by the user, \ourmethod leverages the RAG system to retrieve the latest relevant papers, forming the basis for generating a structured and informed outline. During subsection writing, the system dynamically pulls in new research articles relevant to the specific content under development. This approach ensures that citations are current and the survey content is aligned with the latest developments in the field, significantly enhancing the accuracy and depth of the literature review. \textbf{Muti-LLM-as-judge evaluation}: \ourmethod employs the Multi-LLM-as-Judge strategy, leveraging the LLM-as-Judge method for text evaluation~\cite{zheng2024judging,pandalm2024,yu2024kieval}. This approach generates initial evaluation metrics using multiple large language models, which process a substantial corpus of high-quality surveys. These metrics are refined by human experts to ensure precision and adherence to academic standards. The Multi-LLM-as-Judge method assesses generated content across two main dimensions: (1) Citation Quality, verifying the accuracy and reliability of the information presented, with sub-indicators for Recall and Precision. (2) Content Quality, consisting of Coverage (assessing the extent of topic encapsulation), Structure (evaluating logical organization and coherence), and Relevance (ensuring alignment with the main topic). By utilizing multiple LLMs, this strategy minimizes bias and ensures a balanced and comprehensive assessment, upholding rigorous academic standards.

Extensive experimental results across different survey lengths (8k, 16k, 32k, and 64k tokens) demonstrate that \ourmethod consistently achieves high citation and content quality scores. At 64k tokens, \ourmethod achieves 82.25\% recall and 77.41\% precision in citation quality, outperforming naive RAG-based LLMs (68.79\% recall and 61.97\% precision) and approaching human performance (86.33\% recall and 77.78\% precision). In content quality at 64k tokens, \ourmethod scores 4.73 in coverage, 4.33 in structure, and 4.86 in relevance, closely aligning with human performance (5.00, 4.66, and 5.00 respectively). At shorter lengths (8k, 16k, and 32k tokens), \ourmethod also maintains strong performance across all metrics. Furthermore, the Spearman's rho values indicate a moderate positive correlation between the rankings provided by the LLMs and those given by human experts. The mixture of models achieves the highest correlation at 0.5429, indicating a strong alignment with human preferences. These results reinforce the effectiveness of our multi-LLM scoring mechanism, providing a reliable proxy for human judgment across varying survey lengths.

In conclusion, to the best of our knowledge, \ourmethod is the first system to explore the potential of large model agents in writing extensive academic surveys. It proposes evaluation criteria for surveys that align with human preferences, providing a valuable reference for future related research.

% By open-sourcing \ourmethod with the associated resources at \url{https://github.com/}\wyd{add}, we hope to facilitate further research and inspire new advancements in this area.

\section{Methodology}

\begin{figure}[t!]
\centering
    \includegraphics[width=0.9\textwidth]{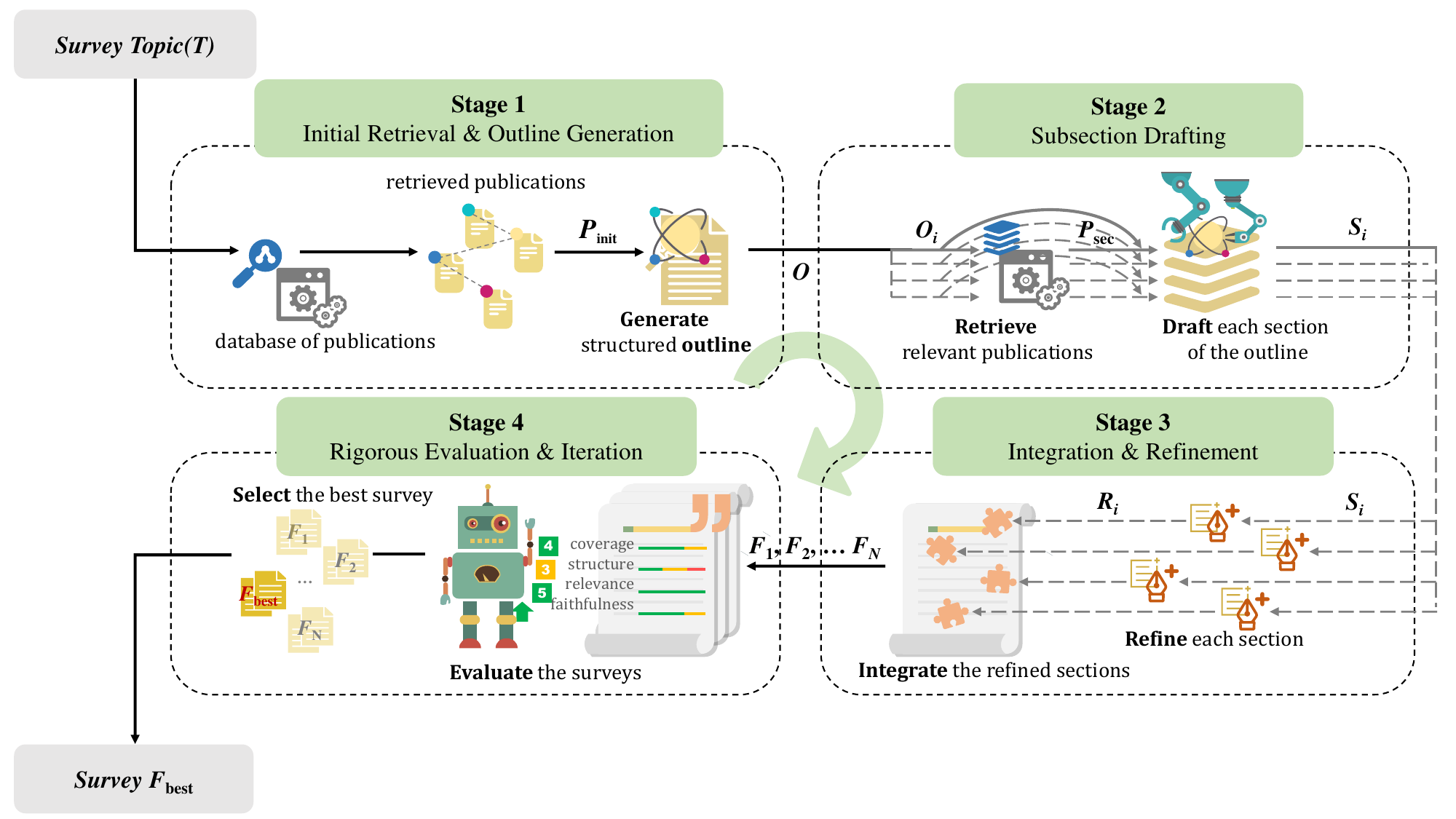}
    \caption{The \ourmethod Pipeline for Generating Comprehensive Surveys.}
    \label{fig-main-fig}

\end{figure}

In this section, we describe the methodology employed by \ourmethod to automate the creation of comprehensive literature surveys. Our approach systematically progresses through four distinct phases---Initial Retrieval and Outline Generation, Subsection Drafting, Integration and Refinement, and Rigorous Evaluation and Iteration. Each phase is meticulously designed to address specific challenges associated with survey creation, thereby enhancing the efficiency and quality of the resulting survey document. The pseudo code of \ourmethod can be found at Algorithm~\ref{alg:autosurvey}.

\scalebox{0.8}{ % 设置缩放比例为0.8，即缩小到原来的80%
\parbox{\linewidth}{ % 使用\parbox以保持宽度一致
\begin{algorithm}[H]
\centering
\caption{\textsc{AutoSurvey}: Automated Survey Creation Using LLMs.}
\label{alg:autosurvey}

\begin{algorithmic}[1]
\STATE \textbf{Input:} Survey topic $T$, publications database $D$
\STATE \textbf{Output:} Final refined and evaluated survey document $F_{best}$

\FOR{each survey generation trial $t = 1$ to $N$}

\STATE \textbf{Phase 1: Initial Retrieval and Outline Generation}
\STATE Retrieve initial pool of publications $P_{\text{init}} \leftarrow \text{Retrieve}(T, D)$
\STATE Generate outline $O \leftarrow \text{Outline}(T, P_{\text{init}})$

\STATE \textbf{Phase 2: Subsection Drafting}
\FOR{each section $O_i$ in $O$ \textbf{in parallel}}
\STATE Retrieve relevant publications $P_{\text{sec}} \leftarrow \text{Retrieve}(O_i, D)$
\STATE Draft subsection $S_i \leftarrow \text{Draft}(O_i, P_{\text{sec}})$
\ENDFOR

\STATE \textbf{Phase 3: Integration and Refinement}

\STATE Refine the merged document to improve coherence $R_i \leftarrow \text{Refine}(S_i)$
\STATE Merge subsection drafts into a single document $F_t \leftarrow \text{Merge}(R_1, R_2, \ldots, R_n)$

\ENDFOR

\STATE \textbf{Phase 4: Rigorous Evaluation and Iteration}
\STATE Evaluate and select the best survey document $F_{\text{best}} \leftarrow \text{Evaluate}({F_1, F_2, \ldots, F_N})$

\STATE \textbf{Return:} Refined and evaluated survey $F_{\text{best}}$
\end{algorithmic}
\end{algorithm}
}
}

\paragraph{Initial Retrieval and Outline Generation}
The process begins with the Initial Retrieval and Outline Generation phase. Utilizing an embedding-based retrieval technique, \ourmethod scans a database of publications to identify papers most pertinent to the specified survey topic \(T\). This phase is crucial for ensuring that the survey is grounded in the most relevant and recent research. The retrieved publications \(P_{\text{init}}\) are then used to generate a structured outline \(O\), which ensures comprehensive coverage of the topic and logical structuring of the survey. To provide more detailed guidance for writing subsections, the outline generation includes not only titles for each subsection but also brief descriptions. These descriptions convey the main idea of each subsection, aiding in the overall clarity and direction of the survey. Given the extensive amount of relevant papers extracted during this stage, the total length of \(P_{\text{init}}\) often exceeds the context window size of the LLM. To address this, papers are randomly divided according to the LLM's context window size, resulting in the creation of multiple outlines. The model then consolidates these outlines to form the final comprehensive outline. Finally, the outline \(O\) of the entire survey is represented as $O = \text{Outline}(T, P_{\text{init}}).$

\paragraph{Subsection Drafting}
With the structured outline in place, the Subsection Drafting phase commences. During this phase, specialized LLMs draft each section of the outline in parallel. This method not only accelerates the drafting process but also ensures detailed and focused content generation for each survey section, adhering to the thematic boundaries established by the outline. When writing the content of each subsection, the sub-outline \(O_i\) of that subsection will be used to retrieve the necessary relevant reference papers \(P_{\text{sec}}\) to provide information that aligns more closely with the main idea of the subsection. During the writing process, the model is required to cite the provided reference papers to support the generated content. The references in the generated content will be extracted and mapped to the corresponding arXiv papers (see Appendix \ref{appendix:implementation} for details). The \(i_{th}\) subsection \(S_i\) can be expressed as: $S_i = \text{Draft}(O_i, P_{\text{sec}}).$

\paragraph{Integration and Refinement}
Following the drafting phase, each section \(S_i\) is individually refined to enhance readability, eliminate redundancies, and ensure a seamless narrative. The refined sections \(R_i\) are then merged into a cohesive document \(F\), which is essential for maintaining a logical flow and coherence throughout the survey. During the refinement process, the model needs to polish each subsection based on the local context (considering the previous and following subsections) to improve readability, eliminate redundancies, and enhance coherency. Additionally, the model is required to check the correctness of the cited references in the content and correct any errors in the citations. This procedure can be represented by:
$F = \text{Merge}(R_1, R_2, \ldots, R_n), \text{where} \ R_i = \text{Refine}(S_i).$

\paragraph{Rigorous Evaluation and Iteration}
The final phase involves a rigorous evaluation and iteration process, where the survey document is assessed through a Multi-LLM-as-Judge strategy. This evaluation critically examines the survey in several aspects. The insights gained from this evaluation are used to guide further refinements, ensuring the survey meets the highest academic standards. The best survey is chosen from \(N\) candidates. The final output of \ourmethod is $ F_{\text{best}} = \text{Evaluate}(\{F_1, F_2, \ldots, F_N\}).$

The methodology outlined here—from initial data retrieval to sophisticated multi-faceted evaluation—ensures that \ourmethod effectively addresses the complexities of survey creation in evolving research fields using advanced LLM technologies.

\section{Experiments}
\paragraph{Setup}

We conduct comprehensive experiments to evaluate the performance of \ourmethod, comparing it against traditional methods for generating survey papers. For the drafting phase of \ourmethod, we utilize Claude-3-Haiku, known for its speed and cost-effectiveness, capable of handling 200K tokens. For evaluations, we employ a combination of GPT-4, Claude-3-Haiku, and Gemini-1.5-Pro\footnote{Specifically, we use gpt-4-0125-preview, claude-3-haiku-20240307 and Gemini-1.5-pro-preview.}. The evaluation covers the following key performance metrics:

\begin{itemize}
\item \textbf{Survey Creation Speed}: To estimate the time it takes for humans to write a document, we use a mathematical model with the following parameters: \(L\) (the length of the document), \(E\) (the number of experts), \(M\) (the writing speed of each expert), \(T_r\) (the preparation time for research and data collection), \(T_w\) (the actual writing time, \(T_w = \frac{L}{E \times M}\)), and \(T_e\) (the editing and revision time, \(T_e = \frac{1}{2} T_w\)). Assuming an ideal situation where \(E = 10\), \(M = 2000\) tokens/hour, \(T_r = 5\) hours, and \(T_e = \frac{1}{2} T_w\), the total time \(Time\) is calculated as:
    \begin{equation}
    Time = T_r + T_w + T_e = T_r + \frac{L}{E \times M} + \frac{1}{2} \times \frac{L}{E \times M}.
    \end{equation}
    For naive RAG-based LLM generation and \ourmethod, we count all the time of API calls. The speed is calculated as \(Speed = \frac{1}{Time (\text{hours})}\).

\item
\textbf{Citation Quality}: Adopted from~\cite{gao2023citation}, this metric assesses the accuracy and relevance of citations in the survey. Assuming a set of claims \(C = \{c_1, c_2, \ldots \}\) extracted from the survey, the metric utilizes an NLI model \(h\) to decide whether a claim \(c_i\) is supported by its references \(\text{Ref}_i = \{r_{i_1}, r_{i_2}, \ldots \}\), where each \(r_{i_k}\) represents one paper cited. \(h(c_i, \text{Ref}_i) = 1\) means that the references can support the claim, and \(h(c_i, \text{Ref}_i) = 0\) otherwise. Refer to Appendix \ref{appendix:evaluation} for more details. Citation quality encompasses two sub-metrics:

\begin{itemize}
    \item \textbf{Citation Recall}: Measures whether all statements in the generated text are fully supported by the cited passages, which is calculated as
    \begin{equation}
        \text{Recall} = \frac{\sum_{i=1}^{|C|} h(c_i, \text{Ref}_i)}{|C|}.
    \end{equation}
    
    \item \textbf{Citation Precision}: Identifies irrelevant citations, ensuring that the provided citations are pertinent and directly support the statements. Before listing the formula for precision, a function \(g\) is defined as \(g(c_i, r_{i_k}) = (h(c_i, \{r_{i_k}\}) = 1) \cup (h(c_i, \text{Ref}_i \setminus \{r_{i_k}\}) = 0)\), which measures whether the paper \(r_{i_k}\) is related to the claim \(c_i\). The precision is
    \begin{equation}
        \text{Precision} = \frac{\sum_{i=1}^{|C|} \sum_{k=1}^{|\text{Ref}_i|} h(c_i, \text{Ref}_i) \cap g(c_i, r_{i_k})}{\sum_{i=1}^{|C|} |\text{Ref}_i|}.
    \end{equation}
\end{itemize}

   \item \textbf{Content Quality}: An overarching metric evaluating the excellence of the written survey, encompassing three sub-indicators. Each sub-indicator is judged by LLMs according to a 5-point rubric, calibrated by human experts to meet academic standards. Note that the detailed scoring criteria are provided in Table~\ref{table: criteria}.
    \begin{itemize}
    \item \textbf{Coverage}: Assesses the extent to which the survey encapsulates all aspects of the topic.
    \item \textbf{Structure}: Evaluates the logical organization and coherence of each section.
    \item \textbf{Relevance}: Measures how well the content aligns with the research topic.
    \end{itemize}
    
\end{itemize}

\begin{table}[ht!]
\centering
\caption{Content Quality Criteria.}
\label{table: criteria}
\tiny
\begin{tabular}{p{1cm} p{12cm}}
\toprule
\textbf{Criteria} & \textbf{Scores} \\
\midrule
\textbf{Coverage} & 

% \textit{Description}: Coverage assesses the extent to which the survey encapsulates all relevant aspects of the topic, ensuring comprehensive discussion on both central and peripheral topics. \\ &

                \textit{Score 1}: The survey has very limited coverage, only touching on a small portion of the topic and lacking discussion on key areas. \\
                  & \textit{Score 2}: The survey covers some parts of the topic but has noticeable omissions, with significant areas either underrepresented or missing. \\
                  & \textit{Score 3}: The survey is generally comprehensive in coverage but still misses a few key points that are not fully discussed. \\
                  & \textit{Score 4}: The survey covers most key areas of the topic comprehensively, with only very minor topics left out. \\
                  & \textit{Score 5}: The survey comprehensively covers all key and peripheral topics, providing detailed discussions and extensive information. \\
\midrule
\textbf{Structure} & 

% \textit{Description}: Structure evaluates the logical organization and coherence of sections and subsections, ensuring that they are logically connected. \\&

\textit{Score 1}: The survey lacks logic, with no clear connections between sections, making it difficult to understand the overall framework. \\
                   & \textit{Score 2}: The survey has weak logical flow with some content arranged in a disordered or unreasonable manner. \\
                   & \textit{Score 3}: The survey has a generally reasonable logical structure, with most content arranged orderly, though some links and transitions could be improved such as repeated subsections. \\
                   & \textit{Score 4}: The survey has good logical consistency, with content well arranged and natural transitions, only slightly rigid in a few parts. \\
                   & \textit{Score 5}: The survey is tightly structured and logically clear, with all sections and content arranged most reasonably, and transitions between adjacent sections smooth without redundancy. \\
\midrule
\textbf{Relevance} & 

% \textit{Description}: Relevance measures how well the content of the survey aligns with the research topic and maintains a clear focus. \\ &

\textit{Score 1}: The content is outdated or unrelated to the field it purports to review, offering no alignment with the topic. \\
                   & \textit{Score 2}: The survey is somewhat on topic but with several digressions; the core subject is evident but not consistently adhered to. \\
                   & \textit{Score 3}: The survey is generally on topic, despite a few unrelated details. \\
                   & \textit{Score 4}: The survey is mostly on topic and focused; the narrative has a consistent relevance to the core subject with infrequent digressions. \\
                   & \textit{Score 5}: The survey is exceptionally focused and entirely on topic; the article is tightly centered on the subject, with every piece of information contributing to a comprehensive understanding of the topic. \\
\bottomrule
\end{tabular}
\end{table}

\paragraph{Baselines} We compare \ourmethod with surveys authored by human experts (collected from Arxiv) and naive RAG-based LLMs across 20 different computer science topics across 20 different topics in the field of LLMs (see Table \ref{table:topics}). For the naive RAG-based LLMs, we begin with a title and a survey length requirement, then iteratively prompt the model to write the content until completion. Note that we also provide the model with the same number of reference papers with \ourmethod.

For \ourmethod, we utilize a corpus of 530,000 computer science papers from arXiv as the retrieval database. During the initial drafting stage, we retrieve 1200 papers relevant to the given topic and split them into several chunks with a window size of 30,000 tokens. The model generates an outline for each chunk and merges these outlines into a final comprehensive outline, using only the abstracts of the papers at this stage. The outline predetermines the number of sections as 8. For subsection drafting, the models generate specific sections using the outline and 60 papers retrieved based on the subsection descriptions, focusing on the main body of each paper (up to the first 1,500 tokens). During the reflection and polishing stage, the same reference papers are provided to the model to ensure consistency and accuracy. The iteration number N is set to 2. Note that human writing surveys used for evaluation are excluded during the retrieval process. For more details of implementations, see Appendix \ref{appendix:implementation}, and the prompts are presented in Appendix \ref{appendix:prompt}.

\paragraph{Main Results}

\begin{table}[ht!]
\centering
\caption{Results of naive RAG-based LLM generation, Human writing and \ourmethod. Note that \ourmethod and naive RAG-based LLM generation both use Claude-haiku as the writer. \textbf{Note that human writing surveys used for evaluation are excluded during the retrieval process.}}
\label{table:main-result}
\begin{adjustbox}{width=0.9\textwidth}
\begin{tabular}{cc|c|cc|cccc}
\toprule
% Survey Length (\#tokens) 
\multirow{2}{*}{Survey Length (\#tokens)}  & \multirow{2}{*}{Methods} & \multirow{2}{*}{Speed} & \multicolumn{2}{c|}{Citation quality} & \multicolumn{4}{c}{Content Quality} \\ 
 & & & Recall & Precision & Coverage & Structure & Relevance & Avg. \\ \midrule
\multirow{3}{*}{8k}    
                       & Human writing     & $0.16$  &  $80.00$ & $87.50$     &  $4.50$   &  $4.16$      &  $5.00$           & $4.52$     \\
                       
                       & Naive RAG-based LLM generation & $79.67$ & $78.14_{\pm 5.23}$ & $71.92_{\pm 6.83}$  & $4.40_{\pm 0.48}$     & $3.86_{\pm 0.71}$      & $4.86_{\pm 0.33}$              & $4.33$\\
                       
                       & AutoSurvey         & $107.00$  & $82.48_{\pm 2.77}$ & $77.42_{\pm 3.28}$ & $4.60_{\pm 0.48}$     & $4.46_{\pm 0.49}$      & $4.8_{\pm 0.39}$                   & $4.61$     \\ \midrule
\multirow{3}{*}{16k} 
                       & Human writing      & $0.14$ & $88.52$ & $79.63$  & $4.66$     & $4.38$      & $5.00$                  & $4.66$      \\
                        & Naive RAG-based LLM generation &$43.41$ & $71.48_{\pm 12.50}$ & $65.31_{\pm 15.36}$ & $4.46_{\pm 0.49}$     & $3.66_{\pm 0.69}$      & $4.73_{\pm 0.44}$                 &$4.23$     \\
                       & AutoSurvey    &   $95.51$  & $81.34_{\pm 3.65}$ & $76.94_{\pm 1.93}$   & $4.66_{\pm 0.47}$     & $4.33_{\pm 0.59}$      & $4.86_{\pm 0.33}$                 & $4.60$      \\ \midrule
\multirow{3}{*}{32k}   
                       & Human writing  & $0.10$ & $88.57$ & $77.14$     & $4.66$     & $4.50$      & $5.00$                  & $4.71$      \\
                       & Naive RAG-based LLM generation &$22.64$ & $79.88_{\pm 4.35}$ & $65.03_{\pm 8.39}$ & $4.41_{\pm 0.64}$     & $3.75_{\pm 0.72}$      & $4.66_{\pm 0.47}$                   &$4.23$      \\
                       & AutoSurvey        & $91.46$  & $83.14_{\pm 2.44}$ & $78.04_{\pm 3.14}$    & $4.73_{\pm 0.44}$     & $4.26_{\pm 0.69}$      & $4.8 _{\pm 0.54}$                 & $4.58$      \\ \midrule
\multirow{3}{*}{64k}  
                       & Human writing & $0.07$  & $86.33$ & $77.78$  & $5.00$     & $4.66$      & $5.00$                  & $4.88$      \\
                        & Naive RAG-based LLM generation & $12.56$ &$68.79_{\pm 11.00}$ & $61.97_{\pm 13.45}$ & $4.4 _{\pm 0.61}$     & $3.66_{\pm 0.47}$      & $4.66_{\pm 0.47}$                & $4.19$      \\
                       & AutoSurvey        & $73.59$  & $82.25_{\pm 3.64}$ & $77.41_{\pm 3.84}$  & $4.73_{\pm 0.44}$     & $4.33_{\pm 0.47}$      & $4.86_{\pm 0.33}$                  & $4.62$    \\ \bottomrule
\end{tabular}
\end{adjustbox}
\end{table}

The results of our experiments comparing human writing, naive RAG-based LLM generation, and \ourmethod for generating academic surveys are summarized in Table~\ref{table:main-result}. The key findings are:

\begin{itemize}
    \item \textit{\ourmethod significantly outperforms both human writing and naive RAG-based LLM generation in terms of speed.} For instance, \ourmethod achieves a speed of 73.59 surveys per hour for a 64k-token survey, compared to 0.07 for human writing and 12.56 for naive RAG-based LLM generation, highlighting the larger gap in speed for longer context generation.
    \item \textit{\ourmethod demonstrates superior citation quality compared to naive RAG-based LLM generation, with performance close to human writing.} For an 8k-token survey, \ourmethod achieves citation recall and precision scores of 82.48 and 77.42, respectively, surpassing naive RAG-based LLM generation (78.14 recall, 71.92 precision). While human writing achieves the best performance, ours is close to human's across different lengths.
    
    \item \textit{\ourmethod excels in content quality, scoring 4.60 on average for a 16k-token survey.} It achieves 4.66 for coverage, 4.33 for structure, and 4.86 for relevance, matching human writing and surpassing naive RAG-based LLM generation.

\end{itemize}

The experiments indicate that \ourmethod provides a balanced trade-off between quality and efficiency. It achieves near-human levels of coverage, relevance, and citation quality while maintaining a significantly lower time cost. While human writing still leads in structure and overall quality, the efficiency and performance of \ourmethod make it a compelling alternative for generating academic surveys. Naive RAG-based LLM, though effective, falls short in several key areas compared to both human writing and \ourmethod, making it the least preferred method among the three for generating high-quality academic surveys, particularly in terms of structure, due to the lack of outline.

\paragraph{Meta Evaluation}
To verify the consistency between our evaluation method and human evaluation, we conduct a meta-evaluation involving human experts and our automated evaluation system. Human experts judge pairs of generated surveys to determine which one is superior. This process, referred to as a "which one is better" game, serves as the golden standard for evaluation. We compare the judgments made by our evaluation method against those made by human experts. Specifically, we provide the experts with the same scoring criteria used in our evaluation for reference. The experts rank the generated 20 surveys, and we compare these rankings with those generated by LLM using Spearman’s rank correlation coefficient to measure consistency between human and LLM evaluations.

\begin{wrapfigure}{r}{0.4\textwidth} % "r" for right, and width of the figure
    \centering
    \includegraphics[width=0.38\textwidth]{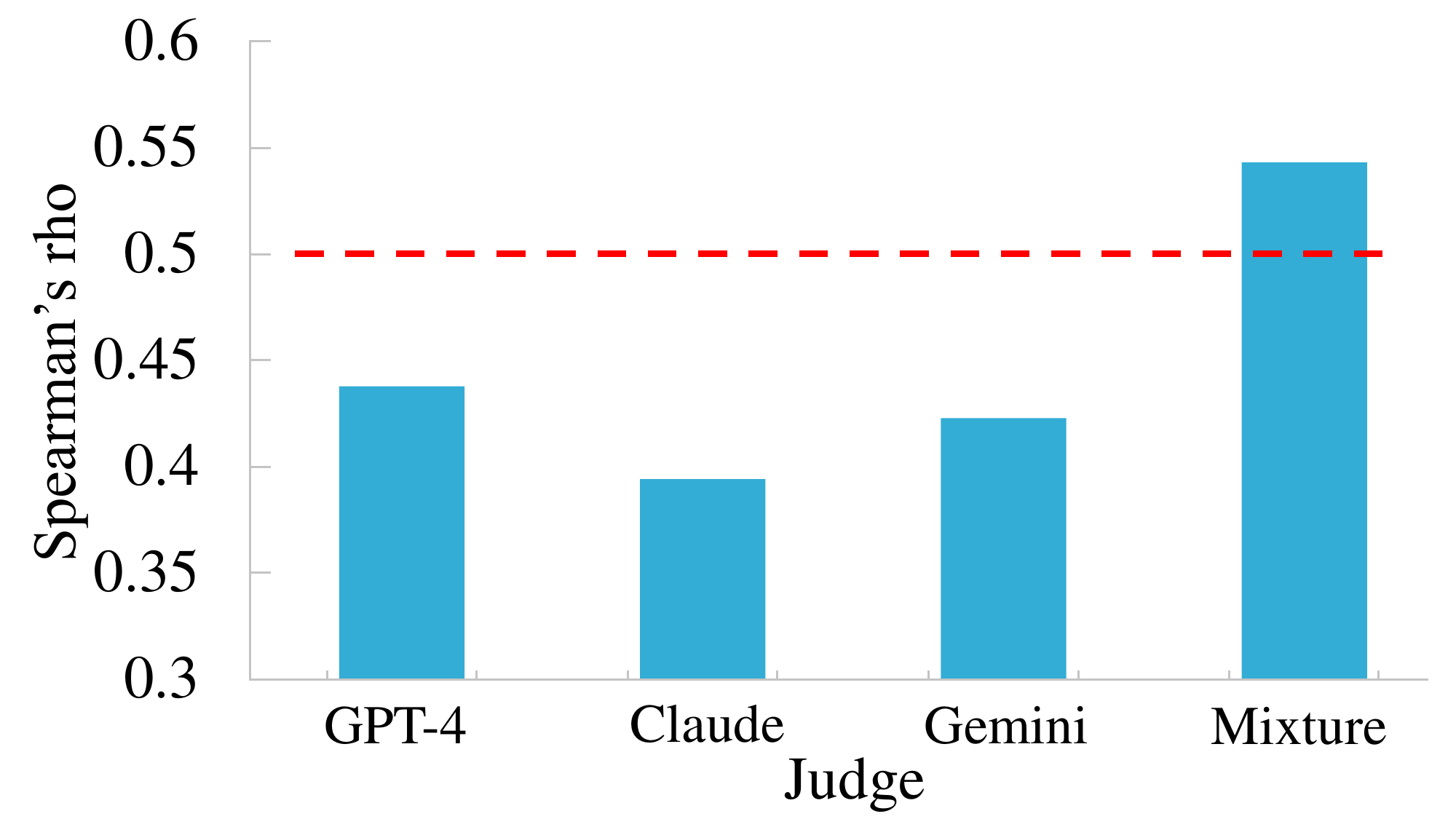}
    \caption{Spearman's rho values indicating the degree of correlation between rankings given by LLMs and human experts. \textit{Note that A value over 0.3 indicates a positive correlation and over 0.5 indicates a strong positive correlation.}}
    \label{figure:spearman}
\end{wrapfigure}

The results of this meta-evaluation are presented in Figure~\ref{figure:spearman}. The table shows the Spearman’s rho values, indicating the degree of correlation between the rankings given by each LLM and the human experts. The Spearman’s rho values indicate a moderate positive correlation between the rankings provided by the LLMs and those given by the human experts, with the mixture of models achieving the highest correlation at 0.5429. These results suggest that our evaluation method aligns well with human preferences, providing a reliable proxy for human judgment.

\paragraph{Ablation study}

To assess the impact of various components on the performance of \ourmethod, we conduct an ablation study by systematically removing key components: the retrieval mechanism, the reflection phase, and iterations. Additionally, we evaluate the influence of using different base LLMs to demonstrate that even with a less optimal LLM like Claude-haiku, \ourmethod's performance remains comparable to human-generated surveys.
% The results are presented in Tables~\ref{table:ablation} and~\ref{table:llm-writer}, and Figure~\ref{figure:iteration}.

\begin{table}[th!]
\centering
\caption{Ablation study results for \ourmethod with different components removed.}
\label{table:ablation}
\begin{adjustbox}{width=0.7\textwidth}
\begin{tabular}{ccc|cccc}
\toprule
% \multirow{2}{*}{Methods  } Methods              & Coverage & Structure & Relevance   & Avg. & Citation quality \\ \midrule

\multirow{2}{*}{Methods} & \multicolumn{2}{c|}{Citation Quality} & \multicolumn{4}{c}{Content Quality} \\
            &    Recall & Precision & Coverage & Structure & Relevance & Avg. \\ \midrule
 
   AutoSurvey &$83.48_{\pm 5.05}$ & $77.15_{\pm 6.05}$ & $4.7_{\pm 0.45}$      & $4.16_{\pm 0.73}$      & $4.93_{\pm 0.30}$     
   & $4.57$
   \\
                        AutoSurvey w/o retrieve    &$60.11_{\pm 6.42}$ & $51.65_{\pm 6.33}$    & $4.51_{\pm 0.49}$     & $4.01_{\pm 0.74}$      & $4.88_{\pm 0.32}$      & $4.44$  
                 \\
                        AutoSurvey w/o reflection     & $83.23_{\pm 3.82}$ & $76.36_{\pm 4.08}$      & $4.76_{\pm 0.42}$     & $4.13_{\pm 0.76}$      & $4.88_{\pm 0.32}$      &  $4.56$

                \\ 
                        % AutoSurvey w/o iteration           & null     & null      & null      & null             & null      \\ 
\bottomrule
\end{tabular}
\end{adjustbox}
\end{table}

Table~\ref{table:ablation} demonstrates that removing the retrieval mechanism significantly degrades citation quality, highlighting its critical role in ensuring accurate and relevant references. The absence of the reflection phase slightly impacts the overall content quality, particularly in structure and coherence.

\begin{table}[th!]
\centering
\caption{Performance of \ourmethod with different base LLM writers.}
\label{table:llm-writer}
\begin{adjustbox}{width=0.6\textwidth}
\begin{tabular}{ccc|cccc}
\toprule
 % Base LLM writer             & Coverage & Structure & Relevance & Avg. & Citation quality \\ \midrule
 \multirow{2}{*}{Base LLM writer} & \multicolumn{2}{c|}{Citation Quality} & \multicolumn{4}{c}{Content Quality} \\ 
& Recall & Precision & Coverage & Structure & Relevance & Avg.  \\ \midrule
   GPT-4 &$80.25_{\pm 4.19}$ & $78.83_{\pm 7.00}$   & $4.8 _{\pm 0.54}$    & $4.46_{\pm 0.49}$      & $4.86_{\pm 0.33}$      & $4.70$             \\
    Claude-haiku     & $82.45_{\pm 2.77}$ & $76.31_{\pm 2.18}$    & $4.66_{\pm 0.47}$     &$4.26_{\pm 0.67}$     & $4.86_{\pm  0.33}$     &$4.58$                   \\
    Gemini-1.5-pro    &$78.13_{\pm 2.39}$ & $71.24_{\pm 3.28}$         & $4.86_{\pm 0.33}$   
    & $4.33_{\pm 0.78}$      & $4.93_{\pm 0.25}$      & $4.69$                   \\ \midrule
     Human    &$85.86$ & $80.51$         & $4.71$   
    & $4.43$      & $5$      & $4.70$  \\
\bottomrule
\end{tabular}
\end{adjustbox}
\end{table}

Table~\ref{table:llm-writer} shows the performance of \ourmethod when using different LLMs as the base writer. The results indicate that all three LLMs (GPT-4, Claude-haiku, and Gemini-1.5-Pro) perform well, with GPT-4 slightly outperforming the others in terms of overall content quality. Importantly, even with the less optimal Claude-haiku, \ourmethod's performance remains competitive with human standards.

\begin{wrapfigure}{r}{0.4\textwidth} % "r" for right, and width of the figure
    \centering
    \includegraphics[width=0.38\textwidth]{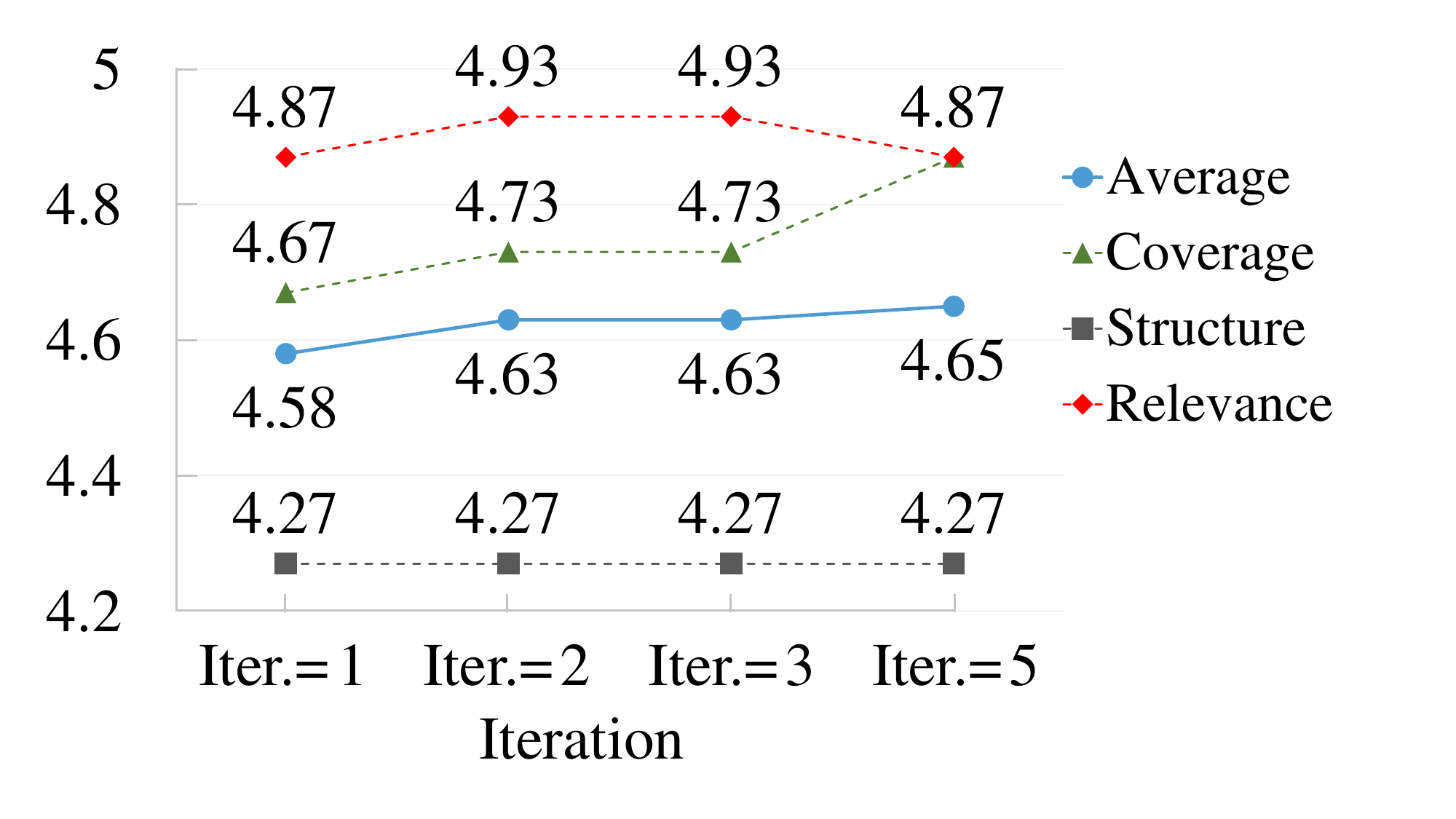}
    \caption{Impact of Iteration on AutoSurvey Performance.}
    \label{figure:iteration}
\end{wrapfigure}

Figure~\ref{figure:iteration} presents the effect of different iteration counts on the performance of \ourmethod. The results show that increasing the number of iterations from 1 to 5 leads to a slight improvement in overall content quality, with diminishing returns after the second iteration.

To assess whether the generated survey can provide useful information to enrich the knowledge, we created 50 multiple-choice questions about 5 topics. These questions primarily involve knowledge related to literature, such as identifying which paper proposed a particular method. We compared the accuracy of the Claude model under the following conditions: (1) directly chooses the answer without providing any reference materials, (2) has access to a 32k survey generated by naive RAG-based LLMs, (3) has access to a 32k survey generated by AutoSurvey, and (4) can refer to 20 papers (30k tokens in total) retrieved using the options provided (Upper-bound, directly retrieving the answers).

\begin{wraptable}{r}{0.3\textwidth}
    \centering
    \caption{Performances given different references.}
    \begin{adjustbox}{width=0.28\textwidth}
    \begin{tabular}{ccc}
        \toprule
        Methods & Accuracy\\
        \midrule
        Direct & $58.40_{\pm{4.96}}$\\
        Naive RAG-based LLMs & $65.20_{\pm{8.06}}$\\
        Upper-bound & $73.60_{\pm{3.44}}$\\
        \ourmethod & $67.60_{\pm{4.96}}$\\
        \bottomrule
    \end{tabular}
    \end{adjustbox}
    \label{tab:choice}
\end{wraptable}

The results are shown in Table \ref{tab:choice} and we find providing topic-related materials can effectively improve the accuracy of answers. Providing option-related papers can be considered an upper bound for performance (73.60\%). AutoSurvey improves accuracy by 9.2\% compared to directly answering and is 2.4\% higher than using naive RAG-based LLM-generated surveys. This demonstrates that our method can effectively provide topic knowledge.

In summary, the ablation study underscores the critical role of the retrieval mechanism and reflection phase in \ourmethod. Furthermore, the performance is influenced by using different LLMs as the base writer and varying the iteration count. Nevertheless, \ourmethod consistently performs well across various configurations, showcasing its robustness and efficiency.

\section{Related Work}
\paragraph{Long-form Text Generation}
The ability to effectively process and generate long-form text is a critical challenge for large language models (LLMs) due to the need to maintain coherence and logical flow over extended passages of text \citep{tan2021long-form,bai2023longbench,dong2023bamboo,li2023LooGLE}. Several works try to address the challenge by directly extending the context window with different Positional Encoding Techniques\citep{shaw2018rpe,wang2019spe}. However, modifying position encoding strategies requires retraining the model, which is costly. Another solution is using memory-augmented techniques. RecurrentGPT \citep{zhou2023recurrentgpt} enables the generation of arbitrarily long texts by simulating the recurrence mechanism of RNNs using natural language prompts to store previous contextual information. Temp-Lora \citep{wang2024templora} enables long text generation by embedding context information into a temporary Lora module updated progressively during generation rather than relying on an extensive context window. These methods effectively establish relationships among tokens and maintain contextual understanding, but still face the issue of long generation times. To further accelerate the generation process, Hierarchical Modeling Techniques have been explored extensively to capture the inherent hierarchical nature of long-form text \citep{fan2018hierarchicalstory,wu2021recursively}. Despite such efficiency, it ignores the long dependency of text and may degrade the content quality \citep{chang2023booookscore}. To tackle the drawbacks, \ourmethod, similarly using a Hierarchical generation paradigm, creates a well-organized outline for guidance and refines the generated content to improve the quality.
\paragraph{Automatic Writing}
Due to the high costs associated with manual writing, automated writing has attracted substantial research interest in recent years. Compared to traditional methods, which primarily focus on training models to generate linguistically coherent text \citep{cho2018traingcohernt, bosselut2018awardcoherent}, the emergency of large language models (LLMs) has opened up new possibilities for automated writing, drawing more attention to broader aspects like faithfulness, logical structure, style, and ethics \citep{zhou2023contextfaithfulness, liu2024probingstructure, zhang2024style, schramowski2022ethics}. For example, Retrieval-Augmented Generation techniques are useful for generating claims with citations \citep{gao2023citation, menick2022teachingcitations}. IRP framework \citep{balepur2023expository} generates expository text by iteratively performing content planning, fact retrieval, and paraphrasing to ensure factuality and stylistic consistency.  Several works focus on the outline creation to improve the structure of generated content. PaperRobot \citep{wang2019paperrobot} incrementally writes key elements to generate a paper abstract. STORM \citep{shao2024assisting} designs a refined outline based on multiple rounds of wiki-page-related Q\&A to facilitate wiki-like article generations. These methods have only been explored in shorter texts (<4k). In contrast, Autosurvey shows its effectiveness in generating long content (64k), with a focus on academic reviews.

\section{Limitation}
In addition to directly using recall and precision to evaluate citations, we also perform a manual analysis, providing a more comprehensive view of the citation quality. We examine 100 unsupported claims and their corresponding references and find that the errors mainly fall into three categories: (1) Misalignment, (2) Misinterpretation, and (3) Overgeneralization. Misalignment occurs when the connection between them is incorrectly made, such as an irrelevant citation. Misinterpretation happens when the claim and source are related, but the claim incorrectly represents the information from the source. Overgeneralization occurs when a claim extends the conclusions of the source material to a broader context than is supported. Among the three types of errors, overgeneralization accounts for the largest proportion (51\%), indicating that LLMs still rely heavily on their parametric knowledge for writing. Misinterpretation has a small proportion (10\%), suggesting that LLMs are capable of understanding the content of the references in most cases, avoiding the creation of claims that significantly deviate from the references.

\textbf{Misalignment (39\%)}: An example is citing the "General Data Protection Regulation (GDPR)" in a context where the referenced paper does not propose GDPR but merely mentions it in the content.

\textbf{Misinterpretation (10\%)}: An example is claiming that "In-context learning allows LLMs to adapt to new tasks by simply conditioning on a few demonstration examples, without the need for any parameter updates or fine-tuning," based on a paper that focuses on meta-out-of-context learning and mentions the limitations of in-context learning.

\textbf{Overgeneralization (51\%)}: An example is that "in-context learning can also benefit from advancements in other learning paradigms, such as multi-task learning," based on a paper that discusses multi-task few-shot learning but does not explicitly address its influence on in-context learning.

Among the three types of errors, overgeneralization accounts for the largest proportion (51\%), indicating that LLMs still rely heavily on their parametric knowledge for writing. Misinterpretation has a small proportion (10\%), suggesting that LLMs are capable of understanding the content of the references in most cases, avoiding the creation of claims that significantly deviate from the references. Additional potential societal impact and ethical considerations are discussed in Appendix \ref{appendix:societal_impact}.

\section{Conclusion}

In this paper, we introduce \ourmethod, a novel methodology leveraging large language models to automate the creation of comprehensive literature surveys. \ourmethod addresses key challenges such as context window limitations and parametric knowledge constraints through a systematic approach involving initial retrieval, outline generation, parallel subsection drafting, integration, and rigorous evaluation. Our experiments show that \ourmethod significantly outperforms naive RAG-based LLM generation and matches human performance in content and citation quality, while also being highly efficient. This advancement offers a scalable and effective solution for synthesizing research literature, providing a valuable tool for researchers in rapidly evolving fields like artificial intelligence.

\bibliography{bibfile}

\begin{thebibliography}{10}

\bibitem{pouyanfar2018survey}
Samira Pouyanfar, Saad Sadiq, Yilin Yan, Haiman Tian, Yudong Tao, Maria~Presa Reyes, Mei-Ling Shyu, Shu-Ching Chen, and Sundaraja~S Iyengar.
\newblock A survey on deep learning: Algorithms, techniques, and applications.
\newblock {\em ACM Computing Surveys (CSUR)}, 51(5):1--36, 2018.

\bibitem{chang2023survey}
Yupeng Chang, Xu~Wang, Jindong Wang, Yuan Wu, Linyi Yang, Kaijie Zhu, Hao Chen, Xiaoyuan Yi, Cunxiang Wang, Yidong Wang, et~al.
\newblock A survey on evaluation of large language models.
\newblock {\em ACM Transactions on Intelligent Systems and Technology}, 2023.

\bibitem{zhao2023survey}
Wayne~Xin Zhao, Kun Zhou, Junyi Li, Tianyi Tang, Xiaolei Wang, Yupeng Hou, Yingqian Min, Beichen Zhang, Junjie Zhang, Zican Dong, et~al.
\newblock A survey of large language models.
\newblock {\em arXiv preprint arXiv:2303.18223}, 2023.

\bibitem{khan2022transformers}
Salman Khan, Muzammal Naseer, Munawar Hayat, Syed~Waqas Zamir, Fahad~Shahbaz Khan, and Mubarak Shah.
\newblock Transformers in vision: A survey.
\newblock {\em ACM computing surveys (CSUR)}, 54(10s):1--41, 2022.

\bibitem{lecun2015deep}
Yann LeCun, Yoshua Bengio, and Geoffrey Hinton.
\newblock Deep learning.
\newblock {\em nature}, 521(7553):436--444, 2015.

\bibitem{goodfellow2016deep}
Ian Goodfellow, Yoshua Bengio, and Aaron Courville.
\newblock {\em Deep learning}.
\newblock MIT press, 2016.

\bibitem{achiam2023gpt}
Josh Achiam, Steven Adler, Sandhini Agarwal, Lama Ahmad, Ilge Akkaya, Florencia~Leoni Aleman, Diogo Almeida, Janko Altenschmidt, Sam Altman, Shyamal Anadkat, et~al.
\newblock Gpt-4 technical report.
\newblock {\em arXiv preprint arXiv:2303.08774}, 2023.

\bibitem{kirillov2023segment}
Alexander Kirillov, Eric Mintun, Nikhila Ravi, Hanzi Mao, Chloe Rolland, Laura Gustafson, Tete Xiao, Spencer Whitehead, Alexander~C Berg, Wan-Yen Lo, et~al.
\newblock Segment anything.
\newblock In {\em Proceedings of the IEEE/CVF International Conference on Computer Vision}, pages 4015--4026, 2023.

\bibitem{touvron2023llama}
Hugo Touvron, Thibaut Lavril, Gautier Izacard, Xavier Martinet, Marie-Anne Lachaux, Timoth{\'e}e Lacroix, Baptiste Rozi{\`e}re, Naman Goyal, Eric Hambro, Faisal Azhar, et~al.
\newblock Llama: Open and efficient foundation language models.
\newblock {\em arXiv preprint arXiv:2302.13971}, 2023.

\bibitem{chen2023extending}
Shouyuan Chen, Sherman Wong, Liangjian Chen, and Yuandong Tian.
\newblock Extending context window of large language models via positional interpolation.
\newblock {\em arXiv preprint arXiv:2306.15595}, 2023.

\bibitem{chen2023longlora}
Yukang Chen, Shengju Qian, Haotian Tang, Xin Lai, Zhijian Liu, Song Han, and Jiaya Jia.
\newblock Longlora: Efficient fine-tuning of long-context large language models.
\newblock In {\em The Twelfth International Conference on Learning Representations}, 2023.

\bibitem{wang2024augmenting}
Weizhi Wang, Li~Dong, Hao Cheng, Xiaodong Liu, Xifeng Yan, Jianfeng Gao, and Furu Wei.
\newblock Augmenting language models with long-term memory.
\newblock {\em Advances in Neural Information Processing Systems}, 36, 2024.

\bibitem{liu2024lost}
Nelson~F Liu, Kevin Lin, John Hewitt, Ashwin Paranjape, Michele Bevilacqua, Fabio Petroni, and Percy Liang.
\newblock Lost in the middle: How language models use long contexts.
\newblock {\em Transactions of the Association for Computational Linguistics}, 12:157--173, 2024.

\bibitem{kaddour2023challenges}
Jean Kaddour, Joshua Harris, Maximilian Mozes, Herbie Bradley, Roberta Raileanu, and Robert McHardy.
\newblock Challenges and applications of large language models.
\newblock {\em arXiv preprint arXiv:2307.10169}, 2023.

\bibitem{shi2023large}
Freda Shi, Xinyun Chen, Kanishka Misra, Nathan Scales, David Dohan, Ed~H Chi, Nathanael Sch{\"a}rli, and Denny Zhou.
\newblock Large language models can be easily distracted by irrelevant context.
\newblock In {\em International Conference on Machine Learning}, pages 31210--31227. PMLR, 2023.

\bibitem{li2023long}
Dacheng Li, Rulin Shao, Anze Xie, Ying Sheng, Lianmin Zheng, Joseph Gonzalez, Ion Stoica, Xuezhe Ma, and Hao Zhang.
\newblock How long can context length of open-source llms truly promise?
\newblock In {\em NeurIPS 2023 Workshop on Instruction Tuning and Instruction Following}, 2023.

\bibitem{li2024long}
Tianle Li, Ge~Zhang, Quy~Duc Do, Xiang Yue, and Wenhu Chen.
\newblock Long-context llms struggle with long in-context learning.
\newblock {\em arXiv preprint arXiv:2404.02060}, 2024.

\bibitem{wang2023surveyfact}
Cunxiang Wang, Xiaoze Liu, Yuanhao Yue, Xiangru Tang, Tianhang Zhang, Cheng Jiayang, Yunzhi Yao, Wenyang Gao, Xuming Hu, Zehan Qi, et~al.
\newblock Survey on factuality in large language models: Knowledge, retrieval and domain-specificity.
\newblock {\em arXiv preprint arXiv:2310.07521}, 2023.

\bibitem{ji2023survey}
Ziwei Ji, Nayeon Lee, Rita Frieske, Tiezheng Yu, Dan Su, Yan Xu, Etsuko Ishii, Ye~Jin Bang, Andrea Madotto, and Pascale Fung.
\newblock Survey of hallucination in natural language generation.
\newblock {\em ACM Computing Surveys}, 55(12):1--38, 2023.

\bibitem{shao2024assisting}
Yijia Shao, Yucheng Jiang, Theodore~A. Kanell, Peter Xu, Omar Khattab, and Monica~S. Lam.
\newblock {Assisting in Writing Wikipedia-like Articles From Scratch with Large Language Models}.
\newblock In {\em Proceedings of the 2024 Conference of the North American Chapter of the Association for Computational Linguistics: Human Language Technologies, Volume 1 (Long and Short Papers)}, 2024.

\bibitem{pandalm2024}
Yidong Wang, Zhuohao Yu, Zhengran Zeng, Linyi Yang, Cunxiang Wang, Hao Chen, Chaoya Jiang, Rui Xie, Jindong Wang, Xing Xie, Wei Ye, Shikun Zhang, and Yue Zhang.
\newblock Pandalm: An automatic evaluation benchmark for llm instruction tuning optimization.
\newblock 2024.

\bibitem{zheng2024judging}
Lianmin Zheng, Wei-Lin Chiang, Ying Sheng, Siyuan Zhuang, Zhanghao Wu, Yonghao Zhuang, Zi~Lin, Zhuohan Li, Dacheng Li, Eric Xing, et~al.
\newblock Judging llm-as-a-judge with mt-bench and chatbot arena.
\newblock {\em Advances in Neural Information Processing Systems}, 36, 2024.

\bibitem{yu2024kieval}
Zhuohao Yu, Chang Gao, Wenjin Yao, Yidong Wang, Wei Ye, Jindong Wang, Xing Xie, Yue Zhang, and Shikun Zhang.
\newblock Kieval: A knowledge-grounded interactive evaluation framework for large language models.
\newblock 2024.

\bibitem{gao2023retrieval}
Yunfan Gao, Yun Xiong, Xinyu Gao, Kangxiang Jia, Jinliu Pan, Yuxi Bi, Yi~Dai, Jiawei Sun, and Haofen Wang.
\newblock Retrieval-augmented generation for large language models: A survey.
\newblock {\em arXiv preprint arXiv:2312.10997}, 2023.

\bibitem{lewis2020retrieval}
Patrick Lewis, Ethan Perez, Aleksandra Piktus, Fabio Petroni, Vladimir Karpukhin, Naman Goyal, Heinrich K{\"u}ttler, Mike Lewis, Wen-tau Yih, Tim Rockt{\"a}schel, et~al.
\newblock Retrieval-augmented generation for knowledge-intensive nlp tasks.
\newblock {\em Advances in Neural Information Processing Systems}, 33:9459--9474, 2020.

\bibitem{jiang2023active}
Zhengbao Jiang, Frank~F Xu, Luyu Gao, Zhiqing Sun, Qian Liu, Jane Dwivedi-Yu, Yiming Yang, Jamie Callan, and Graham Neubig.
\newblock Active retrieval augmented generation.
\newblock In {\em Proceedings of the 2023 Conference on Empirical Methods in Natural Language Processing}, pages 7969--7992, 2023.

\bibitem{gao2023citation}
Tianyu Gao, Howard Yen, Jiatong Yu, and Danqi Chen.
\newblock Enabling large language models to generate text with citations.
\newblock In {\em Proceedings of the 2023 Conference on Empirical Methods in Natural Language Processing}, 2023.

\bibitem{tan2021long-form}
Bowen Tan, Zichao Yang, Maruan Al{-}Shedivat, Eric~P. Xing, and Zhiting Hu.
\newblock Progressive generation of long text with pretrained language models.
\newblock In Kristina Toutanova, Anna Rumshisky, Luke Zettlemoyer, Dilek Hakkani{-}T{\"{u}}r, Iz~Beltagy, Steven Bethard, Ryan Cotterell, Tanmoy Chakraborty, and Yichao Zhou, editors, {\em Proceedings of the 2021 Conference of the North American Chapter of the Association for Computational Linguistics: Human Language Technologies, {NAACL-HLT} 2021, Online, June 6-11, 2021}, pages 4313--4324. Association for Computational Linguistics, 2021.

\bibitem{bai2023longbench}
Yushi Bai, Xin Lv, Jiajie Zhang, Hongchang Lyu, Jiankai Tang, Zhidian Huang, Zhengxiao Du, Xiao Liu, Aohan Zeng, Lei Hou, Yuxiao Dong, Jie Tang, and Juanzi Li.
\newblock Longbench: {A} bilingual, multitask benchmark for long context understanding.
\newblock {\em CoRR}, abs/2308.14508, 2023.

\bibitem{dong2023bamboo}
Zican Dong, Tianyi Tang, Junyi Li, Wayne~Xin Zhao, and Ji{-}Rong Wen.
\newblock {BAMBOO:} {A} comprehensive benchmark for evaluating long text modeling capacities of large language models.
\newblock {\em CoRR}, abs/2309.13345, 2023.

\bibitem{li2023LooGLE}
Jiaqi Li, Mengmeng Wang, Zilong Zheng, and Muhan Zhang.
\newblock Loogle: Can long-context language models understand long contexts?
\newblock {\em CoRR}, abs/2311.04939, 2023.

\bibitem{shaw2018rpe}
Peter Shaw, Jakob Uszkoreit, and Ashish Vaswani.
\newblock Self-attention with relative position representations.
\newblock In Marilyn~A. Walker, Heng Ji, and Amanda Stent, editors, {\em Proceedings of the 2018 Conference of the North American Chapter of the Association for Computational Linguistics: Human Language Technologies, NAACL-HLT, New Orleans, Louisiana, USA, June 1-6, 2018, Volume 2 (Short Papers)}, pages 464--468. Association for Computational Linguistics, 2018.

\bibitem{wang2019spe}
Xing Wang, Zhaopeng Tu, Longyue Wang, and Shuming Shi.
\newblock Self-attention with structural position representations.
\newblock In Kentaro Inui, Jing Jiang, Vincent Ng, and Xiaojun Wan, editors, {\em Proceedings of the 2019 Conference on Empirical Methods in Natural Language Processing and the 9th International Joint Conference on Natural Language Processing, {EMNLP-IJCNLP} 2019, Hong Kong, China, November 3-7, 2019}, pages 1403--1409. Association for Computational Linguistics, 2019.

\bibitem{zhou2023recurrentgpt}
Wangchunshu Zhou, Yuchen~Eleanor Jiang, Peng Cui, Tiannan Wang, Zhenxin Xiao, Yifan Hou, Ryan Cotterell, and Mrinmaya Sachan.
\newblock Recurrentgpt: Interactive generation of (arbitrarily) long text.
\newblock {\em arXiv preprint arXiv:2305.13304}, 2023.

\bibitem{wang2024templora}
Y~Wang, D~Ma, and D~Cai.
\newblock With greater text comes greater necessity: Inference-time training helps long text generation.
\newblock {\em arXiv preprint arXiv:2401.11504}, 2024.

\bibitem{fan2018hierarchicalstory}
Angela Fan, Mike Lewis, and Yann Dauphin.
\newblock Hierarchical neural story generation.
\newblock {\em arXiv preprint arXiv:1805.04833}, 2018.

\bibitem{wu2021recursively}
Jeff Wu, Long Ouyang, Daniel~M Ziegler, Nisan Stiennon, Ryan Lowe, Jan Leike, and Paul Christiano.
\newblock Recursively summarizing books with human feedback.
\newblock {\em arXiv preprint arXiv:2109.10862}, 2021.

\bibitem{chang2023booookscore}
Yapei Chang, Kyle Lo, Tanya Goyal, and Mohit Iyyer.
\newblock Booookscore: A systematic exploration of book-length summarization in the era of llms.
\newblock {\em arXiv preprint arXiv:2310.00785}, 2023.

\bibitem{cho2018traingcohernt}
Woon~Sang Cho, Pengchuan Zhang, Yizhe Zhang, Xiujun Li, Michel Galley, Chris Brockett, Mengdi Wang, and Jianfeng Gao.
\newblock Towards coherent and cohesive long-form text generation.
\newblock {\em arXiv preprint arXiv:1811.00511}, 2018.

\bibitem{bosselut2018awardcoherent}
Antoine Bosselut, Asli Celikyilmaz, Xiaodong He, Jianfeng Gao, Po-Sen Huang, and Yejin Choi.
\newblock Discourse-aware neural rewards for coherent text generation.
\newblock {\em arXiv preprint arXiv:1805.03766}, 2018.

\bibitem{zhou2023contextfaithfulness}
Wenxuan Zhou, Sheng Zhang, Hoifung Poon, and Muhao Chen.
\newblock Context-faithful prompting for large language models.
\newblock {\em arXiv preprint arXiv:2303.11315}, 2023.

\bibitem{liu2024probingstructure}
Jinxin Liu, Shulin Cao, Jiaxin Shi, Tingjian Zhang, Lei Hou, and Juanzi Li.
\newblock Probing structured semantics understanding and generation of language models via question answering.
\newblock {\em arXiv preprint arXiv:2401.05777}, 2024.

\bibitem{zhang2024style}
Chiyu Zhang, Honglong Cai, Yuexin Wu, Le~Hou, Muhammad Abdul-Mageed, et~al.
\newblock Distilling text style transfer with self-explanation from llms.
\newblock {\em arXiv preprint arXiv:2403.01106}, 2024.

\bibitem{schramowski2022ethics}
Patrick Schramowski, Cigdem Turan, Nico Andersen, Constantin~A Rothkopf, and Kristian Kersting.
\newblock Large pre-trained language models contain human-like biases of what is right and wrong to do.
\newblock {\em Nature Machine Intelligence}, 4(3):258--268, 2022.

\bibitem{menick2022teachingcitations}
Jacob Menick, Maja Trebacz, Vladimir Mikulik, John Aslanides, Francis Song, Martin Chadwick, Mia Glaese, Susannah Young, Lucy Campbell-Gillingham, Geoffrey Irving, et~al.
\newblock Teaching language models to support answers with verified quotes.
\newblock {\em arXiv preprint arXiv:2203.11147}, 2022.

\bibitem{balepur2023expository}
Nishant Balepur, Jie Huang, and Kevin Chen-Chuan Chang.
\newblock Expository text generation: Imitate, retrieve, paraphrase.
\newblock {\em arXiv preprint arXiv:2305.03276}, 2023.

\bibitem{wang2019paperrobot}
Qingyun Wang, Lifu Huang, Zhiying Jiang, Kevin Knight, Heng Ji, Mohit Bansal, and Yi~Luan.
\newblock Paperrobot: Incremental draft generation of scientific ideas.
\newblock {\em arXiv preprint arXiv:1905.07870}, 2019.

\bibitem{nussbaum2024nomic}
Zach Nussbaum, John~X. Morris, Brandon Duderstadt, and Andriy Mulyar.
\newblock Nomic embed: Training a reproducible long context text embedder, 2024.

\end{thebibliography}
\bibliographystyle{unsrt}

\appendix

\section{Detail of Topics and Human-writing Surveys}
We select 20 surveys from different topics within the LLM field. During the selection process, we prioritize both the breadth of the topics and the citation count (from google scholar) of the surveys. The basic information of surveys are listed in Table \ref{table:topics}.

\begin{table}[h!]
    \centering
    \tiny
    \caption{Survey Table}
    \begin{tabular}{lll} % 每列对齐方式，这里是左对齐
        \toprule
        \textbf{Topic} & \textbf{Survey Title} & \textbf{Citations} \\ % 表头
        \midrule
         In-context Learning& A survey for in-context learning & 323 \\
         LLMs for Recommendation& A Survey on Large Language Models for Recommendation & 55 \\
         LLM-Generated Texts Detection& A Survey of Detecting LLM-Generated Texts & 42 \\
         Explainability for LLMs& Explainability for Large Language Models & 25 \\
         Evaluation of LLMs& A Survey on Evaluation of Large Language Models &  183\\
         LLMs-based Agents& A Survey on Large Language Model based Autonomous Agents & 101 \\
         LLMs in Medicine& A Survey of Large Language Models in Medicine & 234 \\
         Domain Specialization of LLMs& Domain Specialization as the Key to Make Large Language Models Disruptive & 14 \\
         Challenges of LLMs in Education& Practical and Ethical Challenges of Large Language Models in Education & 53\\
         Alignment of LLMs& Aligning Large Language Models with Human & 53 \\
         ChatGPT& A Survey on ChatGPT and Beyond & 144 \\
         Instruction Tuning for LLMs& Instruction Tuning for Large Language Models & 45 \\
         LLMs for Information Retrieval& Large Language Models for Information Retrieval & 22 \\
         Safety in LLMs& Towards Safer Generative Language Models: Safety Risks, Evaluations, and Improvements & 17 \\
         Chain of Thought& A Survey of Chain of Thought Reasoning & 13 \\
         Hallucination in LLMs& A Survey on Hallucination in Large Language Models & 116 \\
         Bias and Fairness in LLMs& Bias and Fairness in Large Language Models & 12 \\
         Large Multi-Modal Language Models& Large-scale Multi-Modal Pre-trained Models & 61 \\
         Acceleration for LLMs& A Survey on Model Compression and Acceleration for Pretrained Language Models & 22 \\
         LLMs for Software Engineering& Large Language Models for Software Engineering & 49 \\
        \bottomrule
    \end{tabular}
    \label{table:topics}
\end{table}

\section{Details of Implementations}
\label{appendix:implementation}
We adopt nomic-embed-text-v1.5 \citep{nussbaum2024nomic}, a widely used embedding model in RAG applications. To build our database, we store the embeddings of the title and abstract for each paper. Since the context window length is 8k, which is longer than any individual abstract, we embed the raw text directly without chunkings. During generation, related papers are retrieved by the abstract and ranked by their similarity to the query. 
When generating subsection content, the model needs to write the corresponding paper titles where citations are required. After generation, each title will be embedded as a query and be mapped to the closest paper title in our database. This approach allows the LLMs to use their own parameter knowledge to generate citations without references while ensuring the existence of the generated citations. When calling API, we set temperature = 1 and other parameters as default. 
Even with the same parameters, the final length of the generated surveys can vary. Therefore, papers with lengths from 8k to 16k are classified into the 8k category, those from 16k to 32k into the 16k category, and so on.

\section{Details of Evaluation}
\label{appendix:evaluation}
For citation quality, we define a sentence with at least one citation as a claim and extract all the claims from the generated survey. For human evaluations, we invite three PhD students and all of them have experience in writing LLMs-related surveys. We provide them with the same scoring criteria, along with explanations of the specific metrics. They are asked to score based on these criteria, and the final rankings of the generated surveys are determined by the total scores. 
\section{Cost Analysis}
We present the average number of tokens to generate a 32k-tokens survey, along with the cost of using different LLMs in Table \ref{table:price}. 

\begin{table}[h!]
    \centering
    \caption{Cost of \ourmethod}
    \begin{tabular}{lllll} % 每列对齐方式，这里是左对齐
        \toprule
        \textbf{Input tokens} & \textbf{Output tokens}& \textbf{Claude-haiku}&\textbf{Gemini-1.5-pro} & \textbf{GPT-4}\\ % 表头
        \midrule
         3009.7K&112.9K &  0.89\$&11.72\$&33.48\$\\
        \bottomrule
    \end{tabular}
    \label{table:price}
\end{table}

\label{appendix:computational analysis}
\section{Societal Impact and Ethical 
Considerations}
\label{appendix:societal_impact}
By integrating various specialized databases, our approach can generate academic surveys across different fields, potentially filling the gaps in existing reviews. However, as our method relies on the performance of large models, it inevitably contains citation errors. Therefore, the generated survey content is intended for reference only.
All personnel involved in the evaluation process participated voluntarily and received ample compensation. All data used in our experiment is sourced from arXiv and is allowed for non-commercial use.
\section{Prompt used in \ourmethod}
\label{appendix:prompt}
\begin{lstlisting}
ROUGH_OUTLINE_PROMPT = 
    '''
    You want to write a overall and comprehensive academic survey about [TOPIC].
    You are provided with a list of papers related to the topic below:
    ---
    [PAPER LIST]
    ---
    You need to draft a outline based on the given papers.
    The outline should contains a title and several sections.
    Each section follows with a brief sentence to describe what to write in this section.
    The outline is supposed to be comprehensive and contains [SECTION NUM] sections.
    
    Return in the format:
    <format>
    Title: [TITLE OF THE SURVEY]
    Section 1: [NAME OF SECTION 1]
    Description 1: [DESCRIPTION OF SENTCTION 1]
 
    ...
    
    Section K: [NAME OF SECTION K]
    Description K: [DESCRIPTION OF SENTCTION K]
    </format>
    The outline:
    '''

SUBSECTION_OUTLINE_PROMPT = 
    '''
    You want to write a overall survey about [TOPIC].
    You have created a overall outline below:
    ---
    [OVERALL OUTLINE]
    ---
    The outline contains a title and several sections.
    Each section follows with a brief sentence to describe what to write in this section.
    You need to enrich the section [SECTION NAME].
    The description of [SECTION NAME]: [SECTION DESCRIPTION]
    You need to generate the framwork containing several subsections based on the overall outlines.
    Each subsection follows with a brief sentence to describe what to write in this subsection.
    These papers provided for references:
    ---
    [PAPER LIST]
    ---
    Return the outline in the format:
    <format>
    Subsection 1: [NAME OF SUBSECTION 1]
    Description 1: [DESCRIPTION OF SUBSENTCTION 1]
    
    ...
    
    Subsection K: [NAME OF SUBSECTION K]
    Description K: [DESCRIPTION OF SUBSENTCTION K]
    </format>
    Only return the outline without any other informations:
    '''

MERGING_OUTLINE_PROMPT = 
    '''
    You want to write a overall survey about [TOPIC].
    You are provided with a list of outlines as candidates below:
    ---
    [OUTLINE LIST]
    ---
    Each outline contains a title and several sections.
    Each section follows with a brief sentence to describe what to write in this section.
    You need to generate a final outline based on these provided outlines to make the final outline show comprehensive insights of the topic and more logical.
    Return the in the format:
    <format>
    Title: [TITLE OF THE SURVEY]
    Section 1: [NAME OF SECTION 1]
    Description 1: [DESCRIPTION OF SENTCTION 1]
    
    ...
    
    Section K: [NAME OF SECTION K]
    Description K: [DESCRIPTION OF SENTCTION K]
    </format>
    Only return the final outline without any other informations:
    '''

SUBSECTION_WRITING_PROMPT = 
    '''
    You  wants to write a overall and comprehensive survey about [TOPIC].
    You have created a overall outline below:
    ---
    [OVERALL OUTLINE]
    ---
    Below are a list of papers for reference:
    ---
    [PAPER LIST]
    ---
    
    Now you need to write the content for the subsection:
    "[SUBSECTION NAME]".
    The details of what to write in this subsection called [SUBSECTION NAME] is in this descripition:
    ---
    [DESCRIPTION]
    ---
    Here is the requirement you must follow:
    1. The subsection is recommended to contain more than [WORD NUM] words.
    2. When writing sentences that are based on specific papers above, you cite the "paper_title" in a '[]' format to support your content.    
    
    Here's a concise guideline for when to cite papers in a survey:
    ---
    1. Summarizing Research: Cite sources when summarizing the existing literature.
    2. Using Specific Concepts or Data: Provide citations when discussing specific theories, models, or data.
    3. Using Established Methods: Cite the creators of methodologies you employ in your survey.
    4. Supporting Arguments: Cite sources that back up your conclusions and arguments.
    ---
    Only return the content more than [WORD NUM] words you write for the subsection [SUBSECTION NAME] without any other information:
    '''

CITATION_REFLECTION_PROMPT = 
    '''
    You want to write a overall and comprehensive survey about [TOPIC].
    Below are a list of papers for references:
    ---
    [PAPER LIST]
    ---
    You have written a subsection below:
    ---
    [SUBSECTION]
    ---
    The sentences that are based on specific papers above are followed with the citation of "paper_title" in "[]".
    For example 'the emergence of large language models (LLMs) [PaLM: Scaling language modeling with pathways]'
    
    Here's a concise guideline for when to cite papers in a survey:
    ---
    1. Summarizing Research: Cite sources when summarizing the existing literature.
    2. Using Specific Concepts or Data: Provide citations when discussing specific theories, models, or data.
    3. Using Established Methods: Cite the creators of methodologies you employ in your survey.
    4. Supporting Arguments: Cite sources that back up your conclusions and arguments.
    ---
    
    Now you need to check whether the citations of "paper_title" in this subsection is correct.
    Once the citation can not support the sentence you write, correct the paper_title in '[]' or just remove it.

    Do not change any other things except the citations.
    Only return the subsection with correct citations:
    '''

COHERENCY_REFINEMENT_PROMPT = 
    '''
    You want to write a overall and comprehensive survey about [TOPIC].
    
    Now you need to help to refine one of the subsection to improve th ecoherence of your survey.
    
    You are provied with the content of the subsection along with the previous subsections and following subsections.
    
    Previous Subsection:
    --- 
    [PREVIOUS]
    ---
    
    Following Subsection:
    ---
    [FOLLOWING]
    ---
    
    Subsection to Refine: 
    ---
    [SUBSECTION]
    ---
    
    Now refine the subsection to enhance coherence, and ensure that it connects more fluidly with the previous and following subsections. 
    Remember that keep all the essence and core information of the subsection intact. Do not modify any citations in [] following the sentences!!!!
    Only return the whole refined content of the subsection without any other informations:
    '''

\end{lstlisting}
\newpage
\label{appendix:survey_example}
\begin{lstlisting}[language=markdown,style=markdownstyle]
# Comprehensive Survey on Emotion Recognition using Large Language Models

## 1. Introduction to Emotion Recognition and Large Language Models
    Emotion recognition has been a crucial and active research area in the field of affective computing, which aims to enable machines to understand, interpret, and respond to human emotions [1]. Emotions play a fundamental role in human cognition, decision-making, and social interaction [2], and the ability to automatically recognize and interpret emotions has a wide range of applications, including healthcare, education, entertainment, and human-computer interaction [3]. The importance of emotion recognition is evident in various real-world applications. In healthcare, emotion recognition can be used to monitor patient mental health, provide personalized therapy, and improve doctor-patient communication [4]. In education, emotion recognition can help identify student engagement and frustration levels, enabling adaptive learning environments that cater to individual needs [5]. In the entertainment industry, emotion recognition can be used to analyze viewer responses and tailor content to evoke desired emotional responses [6]. Despite the significant benefits of emotion recognition, the field faces several challenges that have hindered its widespread adoption and implementation [7]. One of the primary challenges is the inherent complexity and subjectivity of emotions, which can vary across individuals, cultures, and contexts [8]. Emotions are often expressed through multiple modalities, including facial expressions, vocal cues, body language, and physiological signals, and integrating these diverse sources of information is a significant challenge [9]. Additionally, the availability of high-quality, diverse, and annotated emotion datasets is a persistent challenge in the field [10]. Many existing datasets are limited in size, lack diversity, or have inconsistent or subjective emotion labeling, which can lead to biases and poor generalization of emotion recognition models [11].
    ...
### 1.1 Background on Emotion Recognition

### 1.2 Large Language Models and their Capabilities

### 1.3 Emotion Representation in LLMs

### 1.4 Multimodal Emotion Recognition using LLMs

## 2. Techniques and Approaches for Emotion Recognition using LLMs

### 2.1 Fine-tuning LLMs on Emotion Datasets

### 2.2 LLM-based Prompting Methods for Emotion Recognition

### 2.3 Integrating LLMs with Other Modalities for Multimodal Emotion Recognition

## 3. Enhancing LLM-based Emotion Recognition

### 3.1 Data Augmentation for Improving Emotion Recognition

### 3.2 Prompt Engineering for Emotion Recognition

### 3.3 Integrating External Knowledge for Emotion Recognition

## 4. Challenges, Limitations, and Ethical Considerations

### 4.1 Model Biases and Hallucinations

### 4.2 Interpretability and Explainability

### 4.3 Ethical Considerations

## 5. Applications and Future Directions

### 5.1 Assistive Robotics

### 5.2 Mental Health Assessment

### 5.3 Customer Service and User Experience

### 5.4 Symbolic Reasoning and Long-tailed Emotions

### 5.5 Robust Evaluation Frameworks

## References

[1] Affective Computing for Large-Scale Heterogeneous Multimedia Data  A  Survey
[2] Emotion Recognition in Conversation  Research Challenges, Datasets, and  Recent Advances
[3] A Comprehensive Survey on Affective Computing; Challenges, Trends,  Applications, and Future Directions
[4] Affective Computing for Healthcare  Recent Trends, Applications,  Challenges, and Beyond
[5] Automatic Sensor-free Affect Detection  A Systematic Literature Review
[6] Affective Video Content Analysis  Decade Review and New Perspectives
[7] Emotion Recognition from Multiple Modalities  Fundamentals and  Methodologies
[8] The Ambiguous World of Emotion Representation
[9] Multimodal Affective Analysis Using Hierarchical Attention Strategy with  Word-Level Alignment
[10] Expression, Affect, Action Unit Recognition  Aff-Wild2, Multi-Task  Learning and ArcFace
[11] Feature Dimensionality Reduction for Video Affect Classification  A  Comparative Study

\end{lstlisting}

\end{document}